# High-pressure polymorphism of BaFe$_2$Se$_3$

V. Svitlyk[1*], G. Garbarino[1], A. D. Rosa[1], E. Pomjakushina[2],
A. Krzton-Maziopa[3], K. Conder[2], M. Nunez-Regueiro[4], M. Mezouar[1]

[1]*European Synchrotron Radiation Facility, 38000 Grenoble, France*
[2]*Laboratory for Multiscale Materials Experiments, Paul Scherrer Institute, 5232 Villigen, Switzerland*
[3]*Warsaw University of Technology, Faculty of Chemistry, 00-664 Warsaw, Poland*
[4]*Univ Grenoble Alpes, Inst Neel, CNRS, F-38042 Grenoble, France*
*e-mail: svitlyk@esrf.fr

**Abstract**

BaFe$_2$Se$_3$ is a potential superconductor material exhibiting transition at 11 K and ambient pressure. Here we extended the structural and performed electrical resistivity measurements on this compound up to 51 GPa and 20 GPa, respectively, in order to distinguish if the superconductivity in this sample is intrinsic to the BaFe$_2$Se$_3$ phase or if it is originating from minor FeSe impurities that show a similar superconductive transition temperature. The electrical resistance measurements as a function of pressure show that at 5 GPa the superconducting transition is observed at around 10 K, similar to the one previously observed for this sample at ambient pressure. This indicates that the superconductivity in this sample is intrinsic to the BaFe$_2$Se$_3$ phase and not to FeSe with $T_c > 20$ K at these pressures. Further increase in pressure suppressed the superconductive signal and the sample remained in an insulating state up to the maximum achieved pressure of 20 GPa.

Single-crystal and powder X-ray diffraction measurements revealed two structural transformations in BaFe$_2$Se$_3$: a second order transition above 3.5 GPa from *Pnma* (CsAg$_2$I$_3$-type structure) to *Cmcm* (CsCu$_2$Cl$_3$-type structure) and a first order transformation at 16.6 GPa. Here, $\gamma$-BaFe$_2$Se$_3$ transforms into $\delta$-BaFe$_2$Se$_3$ (*Cmcm*, CsCu$_2$Cl$_3$-type average structure) via a first order phase transition mechanism. This transitions is characterized by a significant shortening of the *b* lattice parameter of $\gamma$-BaFe$_2$Se$_3$ (17%) and accompanied by an anisotropic expansion in the orthogonal *ac* plane at the transition point.

**Introduction**

While the BaFe$_2$Se$_3$ compound was discovered more than 40 years ago [1] only recently it was reported to feature possible superconductivity at 11 K [2]. It remained, however, unclear whether the observed superconducting signal was intrinsic to the $BaFe_2Se_3$ bulk as in some of the studied samples FeSe impurities were present [2-4] with a $T_c$ of 8.5 K [5] which is close to the one observed in $BaFe_2Se_3$. Other independent studies on $BaFe_2Se_3$ [6,7] and $BaFe_{1.79}Se_3$ [8] did not reveal a presence of superconductive signal at ambient pressure.

Room temperature (RT) modification of $BaFe_2Se_3$ crystallizes in a $CsAg_2I_3$-type structure (*Pnma* space group, composed of double chains of edge-sharing $FeSe_4$ tetrahedra separated by Ba atoms) and undergoes an antiferromagnetic ordering at $T_N$ = 240 K [2]. Interestingly, different samples with the same nominal composition of $BaFe_2Se_3$ but with inconsistent cell parameters feature different temperatures of Neel ordering ranging from 140 to 256 K [2-4,6,7,9]. Possible slight deviations from the ideal stoichiometry could be in the origin of these differences. Indeed, the Fe-deficient $BaFe_{1.79}Se_3$ phase does not order magnetically and exhibits a spin glass-like transition at 50 K [8]. Similarly, the recently discovered temperature- and pressure-dependent polymorphs of $BaFe_2Se_3$ [10] are also expected to exhibit differing properties compared to the RT modification of $BaFe_2Se_3$, at least structural ones.

In this work we report a discovery of a new high pressure (HP) modification of $BaFe_2Se_3$, denoted as $\delta$, which brings the total number of known $BaFe_2Se_3$ polymorphs to four. Structural relationships as well as phase transition mechanisms between the different $BaFe_2Se_3$ modifications were studied in details using single crystal and powder synchrotron X-ray diffraction combined with high pressure diamond anvil cell techniques. The *P*-dependent structural data were complemented with electrical resistance measurements and a correlation between the structural and physical properties of $BaFe_2Se_3$ was observed. In addition, we provide evidence that the superconducting transition observed previously at 11 K by [2] on the same sample is intrinsic to the $BaFe_2Se_3$ phase although its superconducting fraction is quite low.

**Experiment**

The studied $BaFe_2Se_3$ sample is from the same growth batch as the one studied in our previous works [2,10]. Single crystals of $BaFe_2Se_3$ were grown using the Bridgman method and the relevant experimental details can be found in Ref. [2]. Both powder and single crystal synchrotron diffraction data were collected at the ID27 High Pressure beamline at the European Synchrotron Radiation Facility (Grenoble, France) using synchrotron radiation with the wavelength of 0.3738 Å of a 3x3 µm$^2$ size.

For the *P*-dependent powder diffraction single crystals of BaFe$_2$Se$_3$ were finely ground and loaded into a high-pressure membrane diamond anvil cell (DAC). The sample was contained in a rhenium gasket with a 125 µm hole mounted on the top of 250 µm diamonds. Helium was used as a pressure transmitting medium (PTM) since it preserves high hydrostaticity up to at least 50 GPa [11]. The measurements were performed up to a pressure of 50.8 GPa with a typical step of 1 GPa (Fig. 1). The pressure was measured using the ruby fluorescence technique [12]. The data were collected using a flat panel PerkinElmer detector and integrated with a PyFAI module [13] as included in the Dioptas software [14].

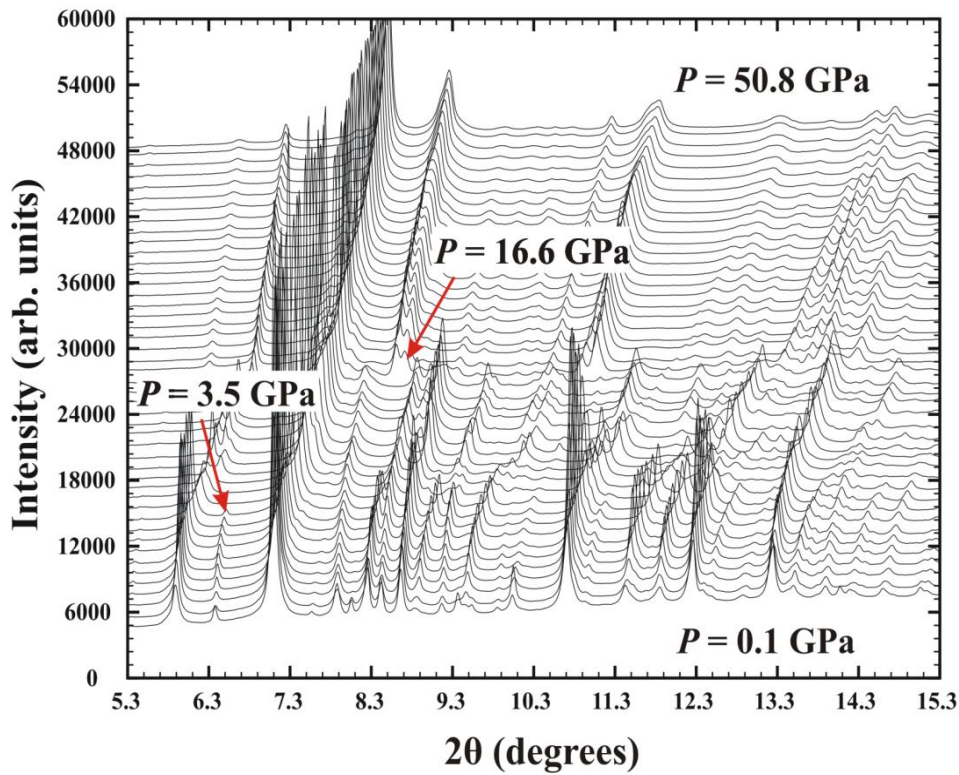

Figure 1. *P*-dependent powder synchrotron radiation diffraction patterns of BaFe$_2$Se$_3$ indicating structural transitions around 3.5 and 16.6 GPa.

For the *P*-dependent single crystal diffraction studies a crystal of BaFe$_2$Se$_3$ of a plate shape with the dimensions of 3x5x15 µm was mounted into a high-pressure membrane DAC with 300 µm size diamonds. The BaFe$_2$Se$_3$ crystal and ruby spheres were contained together with helium gas as a PTM within a 150 µm hole in a rhenium gasket. Measurements were performed at the pressures of 0.2, 9.0 and 21.4 GPa to study all the structural modifications visible on the powder diffraction data (Fig. 1). During the data collection single crystal of BaFe$_2$Se$_3$ was rotated by 60 degrees with the integration steps of 1 degree. The intensities were registered using a MarCCD165 detector and processed with the CryAlisPro software package [15]. Structural solutions and refinements have been performed using the SHELXS and SHELXL programs, respectively [16].

*P*-dependent electrical resistance measurements were performed at the Institut Néel (Grenoble, France) on monocrystalline $BaFe_2Se_3$. A standard four wire resistance method was used [17] in a Bridgman type configuration with pyrophilite gasket and enstatite as PTM. The resistance measurements were performed along the 010 crystallographic direction ($\alpha$-$BaFe_2Se_3$ modification setting) in a pressure range of 1-20 GPa with a typical step of 1.5 GPa. Temperature dependences of the resistance at fixed pressures were collected using a home-made apparatus allowing to measure data on cooling down to 4 K and heating up to 300 K.

**Results and discussion**

From the *P*-dependent powder diffraction data (Fig. 1) the $\alpha$- to $\gamma$-phase transition in $BaFe_2Se_3$, which corresponds to the alignment of the $FeSe_4$ tetrahedra with a corresponding *Pnma* – *Cmcm* symmetry change [10], is triggered at the pressure of 3.5 GPa and is completed at around 4.4 GPa. The analogous transition was originally observed by us at the pressure of 5.5 GPa and was concluded to be of a second order [10]. Indeed, the behaviour of the associate order parameter from the current experiment confirms the second order nature of the transition (Fig. 2, order parameter for the $\alpha$- to $\gamma$-phase transformation expressed through the intensity of the 112 reflection vs. pressure is shown). Consequently, the observed difference of about 2 GPa cannot originate from a structural hysteresis which can be observed only for first-order transformations. The difference could stem from the different hydrostatic properties of the employed PTMs [18]: silicon oil in the original study vs. helium in the current work. Indeed, as it was shown on the example of the related alkali-intercalated FeSe-based superconductors [19], the nature of PTM can influence effective structural and physical response of crystalline solids as a function of pressure.

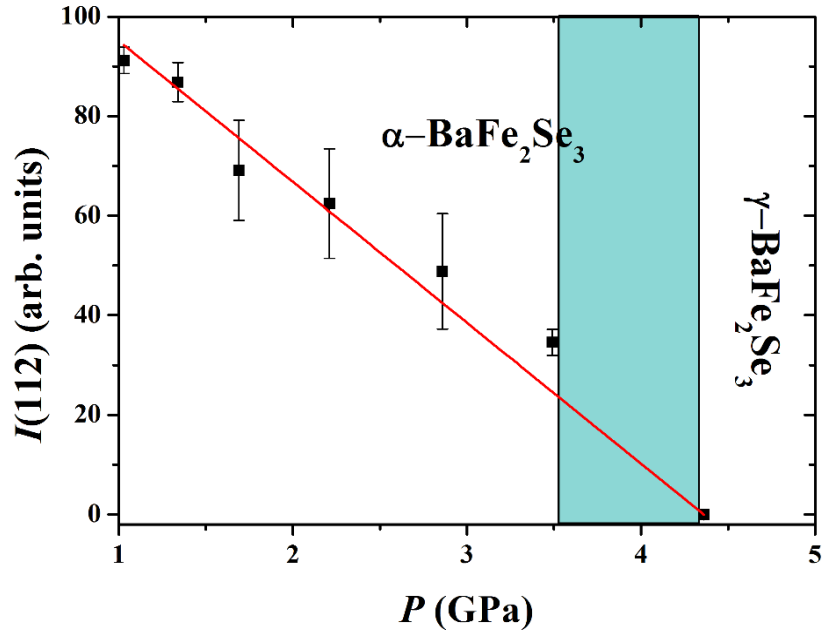

Figure 2. Intensity decrease of the 112 reflection as a function of pressure. The continuous decrease of the reflection intensity is delineated by the red line and indicates a second order α-γ phase transformation in BaFe$_2$Se$_3$. The blue shaded area represents a possible transition region. Intensities and the corresponding error bars have been obtained from the pseudo-Voight fitting of the experimental (112) peaks.

The γ-BaFe$_2$Se$_3$ phase exists up to the pressure of 16.6 GPa, where it undergoes a structural transformation. The structure of the new BaFe$_2$Se$_3$ high pressure modification, which we will denote as σ, was solved from the single crystal data collected at 21.4 GPa. It belongs to the same CsCu$_2$Cl$_3$-type structure (*Cmcm* space group) as the parent γ-BaFe$_2$Se$_3$ modification, but features a collapse in the *b* structural parameter (Fig. 3) accompanied by an anisotropic expansion in the orthogonal *ac* plane. The γ- to σ-phase transition is anti-isostructural and is characterized by an exchange of the *a* and *b* axis of the parent γ-modification with a corresponding transformation matrix equal to 0 -1 0 1 0 0 0 0 1. The experimental atomic parameters obtained from the single crystal refinement are listed in the Table 1.

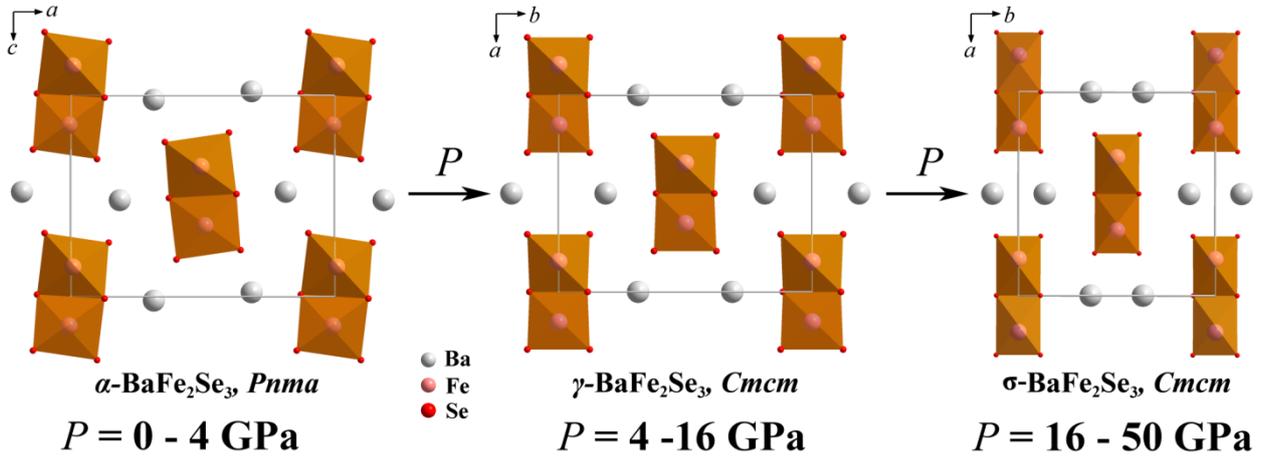

Figure 3. The series of α-γ-σ phase transitions in BaFe$_2$Se$_3$.

Table 1. Atomic parameters for the σ polymorph of BaFe$_2$Se$_3$ at 21.4 GPa from single crystal diffraction (*Cmcm* space group, $a = 9.10(5)$, $b = 8.728(5)$, $c = 5.198(2)$ Å, $R_1 = 11.43$ for the 35 unique reflections with the $I > 4\sigma(I)$ in the ADPs setting).

| Atom  | site | x         | y         | z   | $U_{iso}$ |
|-------|------|-----------|-----------|-----|-----------|
| Ba    | 4c   | 1/2       | 0.1818(2) | 1/4 | 0.05(2)   |
| Se(1) | 4c   | 1/2       | 0.6257(3) | 1/4 | 0.06(3)   |
| Se(2) | 8g   | 0.2067(2) | 0.3780(2) | 1/4 | 0.03(2)   |
| Fe    | 8e   | 0.3545(3) | 1/2       | 0   | 0.09(4)   |

The equivalent (α modification setting) series of the *hk*0 slices illustrate the α *Pnma* to γ *Cmca* transformation with a following change in the orientation matrix for the σ modification (Fig. 4). The nonzero slices of the σ *Cmca* structure reveal features which are not present in the γ *Cmca* modification of BaFe$_2$Se$_3$. Firstly, a series of weak superstructure reflections can be observed at $q = $ ½,½ in the *hk*2 layer (Fig. 5, right, pointed by black arrows). Structural origin of the observed superstructure reflections could not be reliably deduced from the experimental single crystal data due to their weak intensities. These features could stem from short/medium-range order correlations along the *c* direction of σ-BaFe$_2$Se$_3$, possibly related to further reorganization of the FeSe$_4$ double chains.

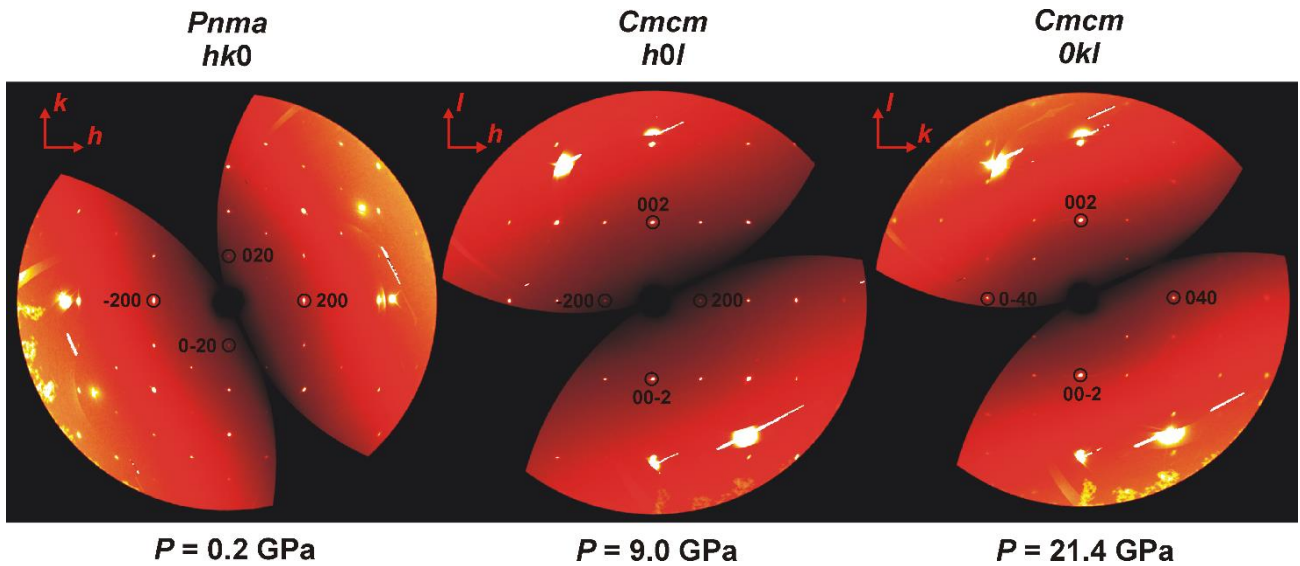

Figure 4. Series of hk0 (left), h0l (center) and 0kl (right) slices of the reciprocal space of the α, γ and σ modifications of $BaFe_2Se_3$, respectively.

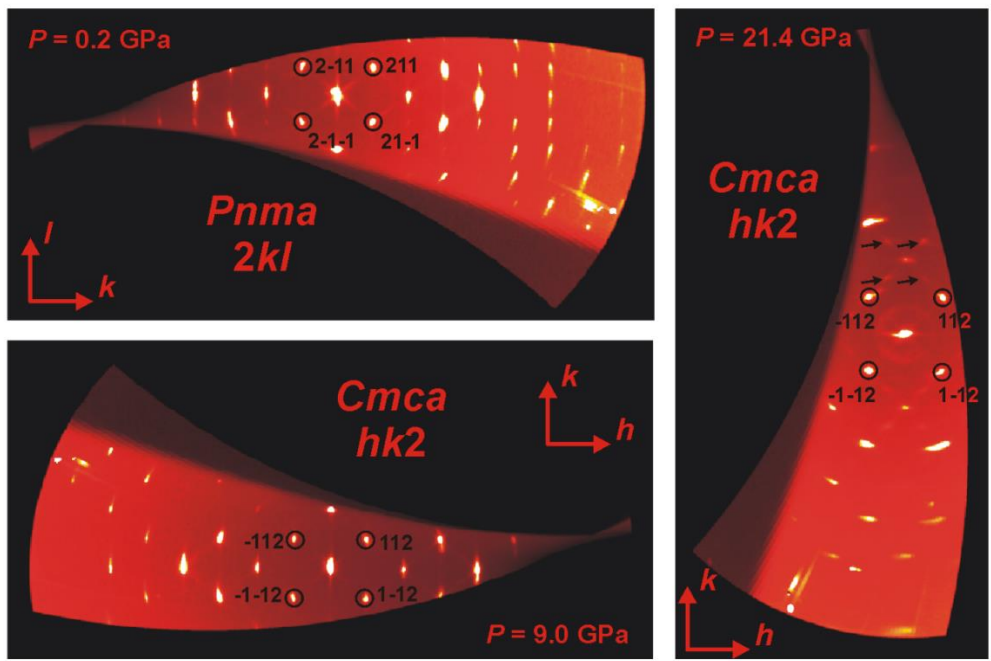

Figure 5. Series of 2kl (top), hk2 (bottom) and hk2 (right) slices of the reciprocal space of the α, γ and σ modifications of $BaFe_2Se_3$, respectively.

Another interesting feature of the σ modification of $BaFe_2Se_3$ is a circle-shaped continuous diffuse scattering visible around the zero order reflection in the hk2 plane (Fig. 5, right). In crystalline solids diffuse scattering can have either atomic or electronic origin. In the first case diffuse scattering can originate from occupational or substitutional atomic correlations [20,21] or a presence of stacking faults [22-24]. Correlated lattice dynamics also can produce quite complex diffuse scattering patterns [25-27]. For the case of σ-BaFe$_2$Se$_3$ occupational and substitutional correlations can be ruled out since no vacancies or atomic mixing is observed in this phase, i.e. it is fully stoichiometric [2,10]. Stacking faults cannot be considered as a possible mechanism as well since this scenario is valid only for layered compounds. Similarly, in crystalline solids application of high pressure is not expected to induce correlated lattice dynamics. Finally, electronically induced diffuse scattering can originate from a possible Fermi surface nesting, as was recently observed for the ThCr$_2$Si$_2$-type compounds, including the BaFe$_2$As$_2$ superconductor [28,29]. A possible electronic contribution to the diffuse scattering of σ-BaFe$_2$Se$_3$ is yet to be established with a help of *ab initio* calculations. The main conclusion to be drawn at this stage is that the CsCu$_2$Cl$_3$-type structure (Table 1) should be considered as an average one for the σ-BaFe$_2$Se$_3$ modification and this structural complexity is expected to influence the observed physical properties.

A detailed behaviour of the unit cell parameters of BaFe$_2$Se$_3$ as a function of pressure was extracted from the powder diffraction data (Fig. 1). As was noted previously, the *b* structural parameter collapses with pressure and the corresponding relative decrease at the γ- to σ-phase transition point is equal to 17% (Fig. 6, left). The *a* parameter increases by 8% at 16 GPa and nearly reaches its original value at ambient conditions (Fig. 6, right).

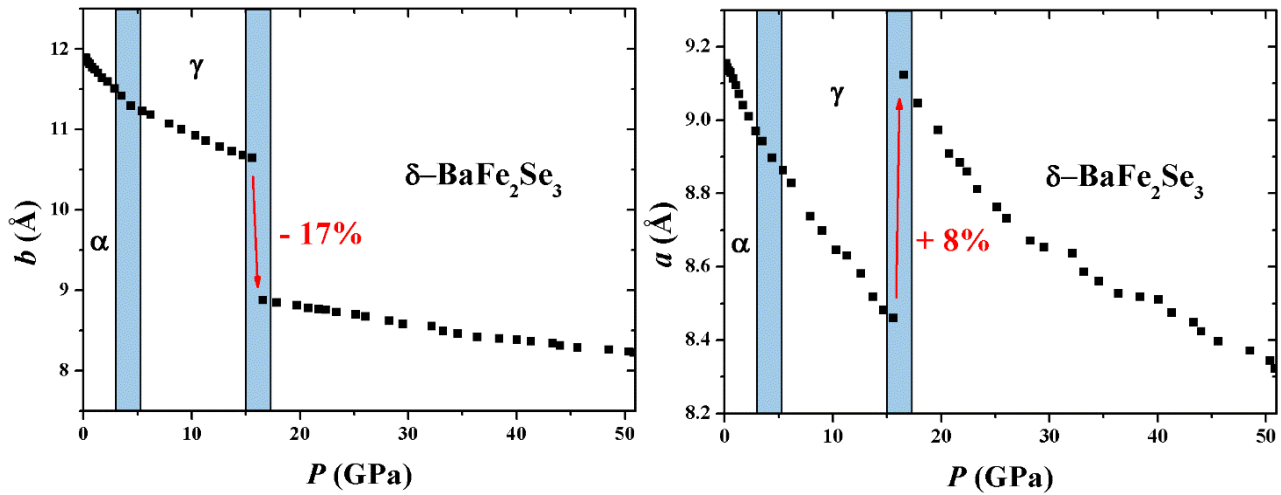

Figure 6. Behaviour of the *b* (left) and *a* (right) structural parameters of BaFe$_2$Se$_3$ as a function of pressure.

Similarly to the *a* lattice parameter, the *c* parameter increases at the γ to σ transition point but only by 3% (Fig. 7, left). Behaviour of the unit cell volume at 16 GPa is dominated by the *b* lattice parameter and, consequently, exhibits a decrease of 8% (Fig. 7, right). The corresponding experimental coefficients of the 3rd order Birch-Murnaghan equation of state (EOS) (Eq. 1, experimental fits are shown as red solid lines on the Fig. 7, right) for the α, γ and σ modifications of BaFe$_2$Se$_3$ are listed in the Table 2. As expected, the high-pressure γ-and σ-phases are less compressible compared to the low-pressure α modification.

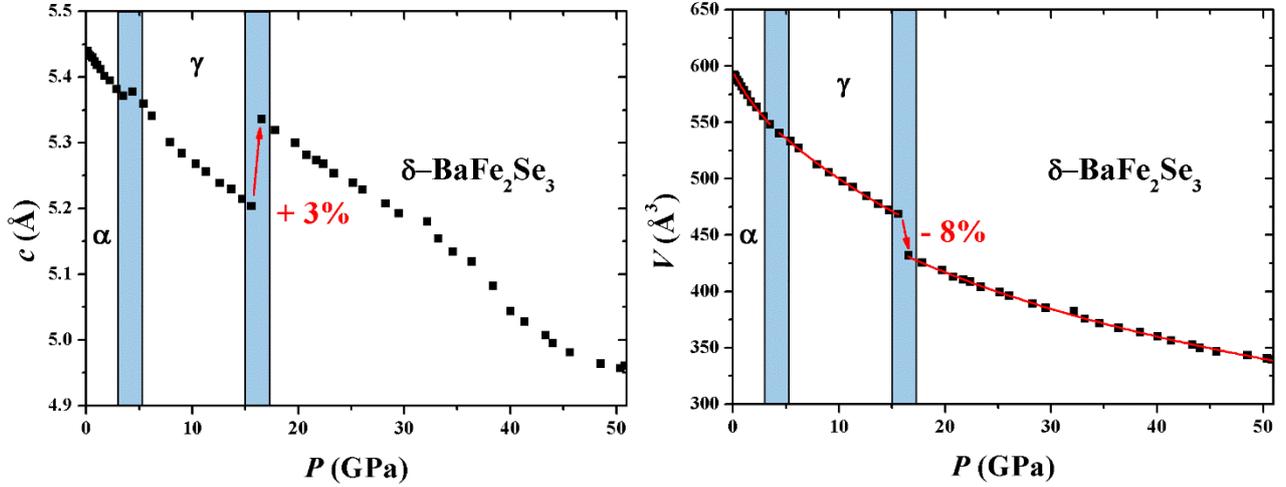

Figure 7. *P*-dependent behaviour of the *c* (left) and *V* (right) structural parameters of BaFe$_2$Se$_3$. The experimental fits of the 3rd order Birch-Murnaghan EOS are shown as red solid lines.

$$P(V) = \frac{3B_0}{2}\left[\left(\frac{V_0}{V}\right)^{\frac{7}{3}} - \left(\frac{V_0}{V}\right)^{\frac{5}{3}}\right]\left\{1 + \frac{3}{4}(B_0' - 4)\left[\left(\frac{V_0}{V}\right)^{\frac{2}{3}} - 1\right]\right\} \qquad \text{Eq. 1}$$

Table 2. Coefficients $V_0$ (zero pressure volumes), $B_0$ (bulk moduli) and $B_0'$ (their first pressure derivatives) of the 3rd order Birch-Murnaghan EOS for the α, γ and σ modifications of BaFe$_2$Se$_3$.

| Phase | $V_0$, Å$^3$ | $B_0$, GPa | $B_0'$, GPa |
|---|---|---|---|
| α-BaFe$_2$Se$_3$ | 594.7(5) | 33(2) | 7(1) |
| γ-BaFe$_2$Se$_3$ | 583(3) | 52(5) | 2.9(5) |
| σ-BaFe$_2$Se$_3$ | 544(15) | 48(9) | 3.5(3) |

The α- to γ-phase transition, which is of second order nature, triggers only minor anomalies in the structural parameters around the transition pressure of 3.5 GPa (Fig. 6 and 7). Contrary, the γ - to σ-phase transformation at around 16 GPa is accompanied by an abrupt structural changes (Fig. 6 and 7), which are characteristic of a first-order phase transition. Indeed, coexistence of the γ-400 and σ-040 reflections of the corresponding γ and σ phases in a 1 GPa pressure range could be observed only for first-order structural transformations (Fig. 8, corresponding series of single crystal diffraction images collected upon increase of pressure are shown).

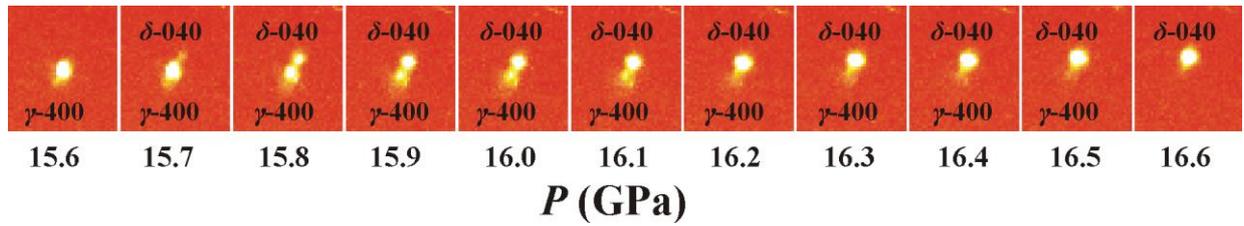

Figure 8. Sections of the *P*-dependent single crystal diffraction images of $BaFe_2Se_3$ showing coexistence of the *γ*-400 and *σ*-040 reflections at the *γ*-*σ* transition point.

Remarkably, despite a strong first-order character of the *γ*-*σ* transition which is accompanied by abrupt changes in structural parameters, the sample remained in a monocrystalline form after the transformation. The origin of this stability can be linked to a morphological similarity of the *γ*- and *σ*-modifications of $BaFe_2Se_3$: both structures are composed of chains of edge-sharing $FeSe_4$ tetrahedra separated by Ba atoms which results in a conservation of the domain structure of the sample upon a transition. Indeed, sizes and shapes of the experimental Bragg reflections of the *γ*- and *σ*-modifications of $BaFe_2Se_3$, which are directly related to the domain structure of the studied $BaFe_2Se_3$ crystal, are identical throughout all the studied pressure range (Figs. 5 and 8).

Taking into account the *σ*-$BaFe_2Se_3$ modification, a phase diagram of the system, as adapted from the ref [10], can be presented in a following way (Fig. 9).

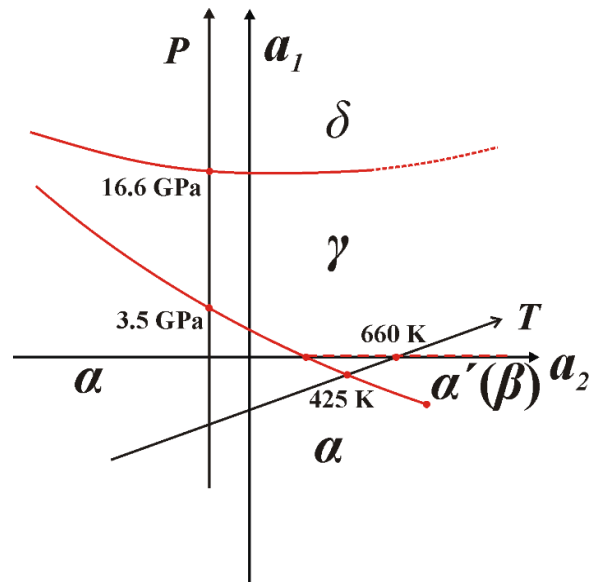

Figure 9. Schematic phase diagram of the $BaFe_2Se_3$ system in the *P*-*T* and *a*1-*a*2 coordinates. The *a*1 and *a*2 symbols are the coefficients of the corresponding expansion of Landau potential [10].

The high-temperature part of the γ-σ phase boundary (dashed line) goes up intentionally since we do not believe that the σ-BaFe$_2$Se$_3$ modification could be achieved only by the application of temperature: the γ-BaFe$_2$Se$_3$ phase likely melts before reaching the γ-σ transition point. An additional experimental mapping of the *P-T* space is needed to pinpoint the real shape of the γ-σ phase boundary of the system.

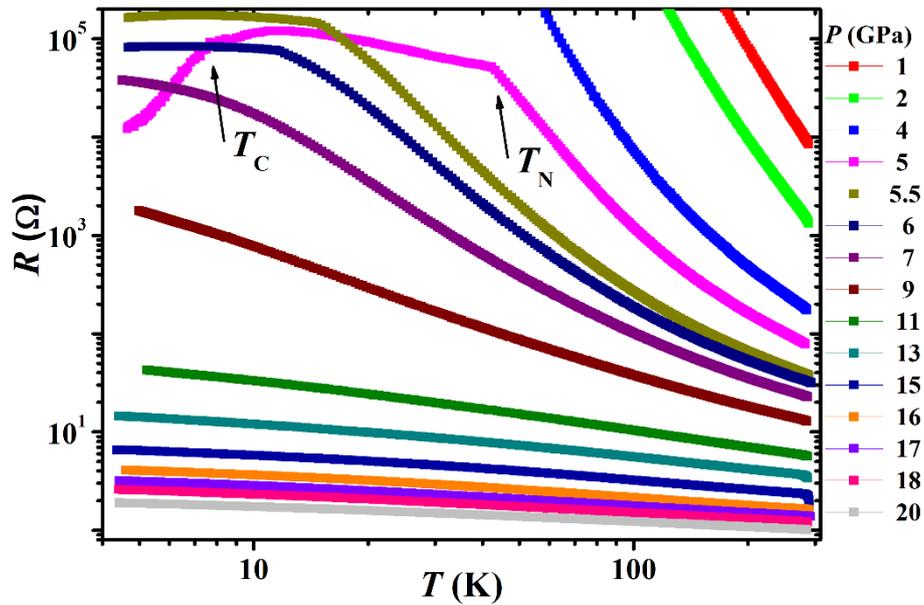

Figure 10. Electrical resistance of BaFe$_2$Se$_3$ (Ω) as a function of pressure (GPa) and temperature (K).

As seen from the experimental *P*-dependent electrical resistance measurements (Fig. 10) the overall response of BaFe$_2$Se$_3$ is dominated by insulating properties. At low pressures (below 5 GPa, Fig. 10), the insulting behaviour of the sample and the experimental configuration does not allow to measure above the maximum measurable resistance. For pressure above 5 GPa, due to the pressure induced reduction of the resistance, we were able to follow the temperature evolution down to 4 K. At 5 GPa two anomalies are evident on the curve, one at 45 K and the other one at around 10 K. We can associate the first one with the antiferromagnetic transition which occurs at 240 K at ambient pressure [2] and reduced to 45 K by application of pressure. This scenario would imply a rapid decrease of the T$_N$ of BaFe$_2$Se$_3$ as a function of pressure. Indeed, the corresponding anomaly on the resistance curve at the next pressure point of 5.5 GPa is shifted down to 12 K (Fig. 10) yielding a change rate of ca. -41 K/GPa in this pressure range. Following increase in pressure further shifts this anomaly to lower temperatures with its complete suppression for pressure above 6 GPa.

Coming back to the second anomaly at around 10 K at 5 GPa mentioned above, it can be associated with the superconducting transition observed for this sample at 11 K at ambient pressure and the corresponding superconducting fraction is low. Indeed, as was shown in our original paper, the superconducting fraction of the studied $BaFe_2Se_3$ sample is close to 1%, as was concluded from the magnetic susceptibility measurements [2]. The superconducting transition temperature close to 10 K at 5 GPa excludes the FeSe phase as a possible source of superconductivity in this sample since at this pressure its critical temperature is higher than 20 K [30-32]. Interestingly, this anomaly is completely supressed at 5.5 GPa and the tendency toward the metallic state is preserved. It is important to stress that in this pressure range the α–γ transition occurs which indicates that this structural change affects the superconducting transition in a drastic way.

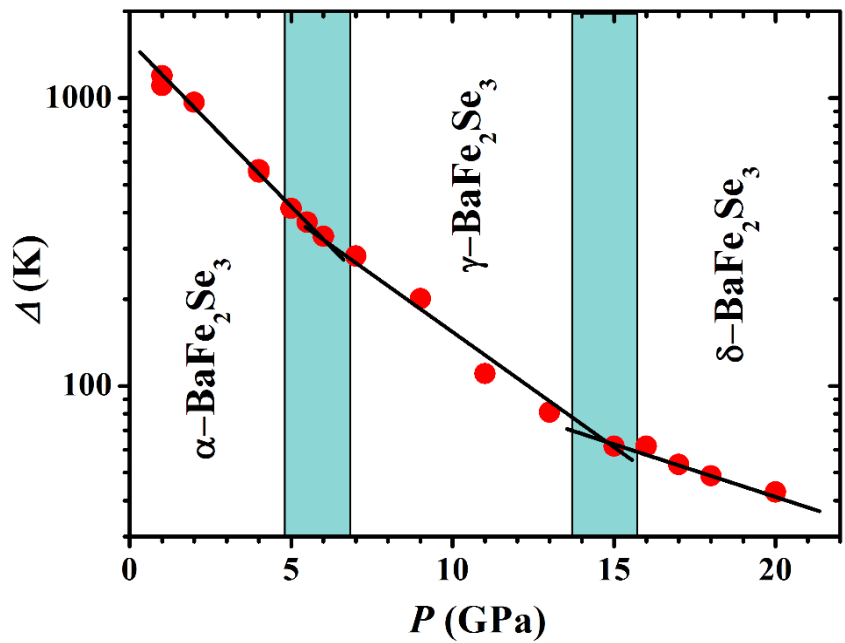

Figure 11. Activation energy, $\Delta$, as a function of pressure for $BaFe_2Se_3$ featuring anomalies around 6 and 15 GPa.

While increase in pressure progressively reduces the electrical resistance of $BaFe_2Se_3$ due to a $P$-induced reduction of the bandgap, a metallic behaviour is not reached up to the maximum applied pressure of 20 GPa (Fig. 10). In order to get more information about the evolution of the insulating state with pressure, we obtained the activation energy, $\Delta$, by fitting the resistance curve with the equation $\rho = \rho_0 \cdot e^{\Delta/kT}$ in the temperature range 90K – 260 K. On Figure 11 we present the pressure evolution of the activation energy, $\Delta$, where we can clearly observe two anomalies at around 6 and 15 GPa, which are in good agreement with the α–γ and γ–σ structural phase transitions observed from diffraction studies and the suppression of superconductivity around 5.5 GPa.

**Conclusions**

High pressure single crystal and powder synchrotron X-ray diffraction allowed to discover another structural modification of BaFe$_2$Se$_3$ at 16.6 GPa, denoted as σ, thus bringing the total number of the known polymorphs in the system to four. The σ-BaFe$_2$Se$_3$ modification is formed due to a collapse of the $b$ structural parameter of the parent γ-BaFe$_2$Se$_3$ phase accompanied by an anisotropic increase in the orthogonal $ac$ plane. Remarkably, at the γ-σ transition point the $a$ structural parameter of σ-BaFe$_2$Se$_3$ nearly reaches its original value at ambient conditions. The σ-BaFe$_2$Se$_3$ phase crystallizes in the same CsCu$_2$Cl$_3$-type structure as the parent γ-BaFe$_2$Se$_3$ modification but with a swap of the $a$ and $b$ axis of the parent γ phase. However, the observed additional fine features in the reciprocal space of σ-BaFe$_2$Se$_3$, namely diffuse scattering and weak superstructure reflections, indicates presence of additional correlations in σ-BaFe$_2$Se$_3$ as compared to the parent γ-BaFe$_2$Se$_3$ phase. These phenomena are yet to be modelled with *ab initio* methods based on the average structure presented in this work.

In the studied BaFe$_2$Se$_3$ sample superconductive signal at 5 GPa was observed around 10 K which indicates that: i) this response is intrinsic to the system and cannot stem from the FeSe impurity phase ii) the $T_c$ of BaFe$_2$Se$_3$ is virtually pressure-independent up to 5 GPa. The observed superconductivity is supressed already at 5.5 GPa and the sample exhibits solely insulating behaviour. While further increase in pressure gradually supresses the insulating state of BaFe$_2$Se$_3$, a fully metallic behaviour is not reached up to 20 GPa. Nevertheless, anomalies in the resistance induced by the corresponding $P$-dependent structural transitions are visible, although rather subtle since the global structural arrangement is preserved throughout all the studied pressure range.

The superconducting volume fraction of the studied BaFe$_2$Se$_3$ sample is quite low and further targeted modifications of BaFe$_2$Se$_3$ are needed to improve it. These may include specific annealing conditions and precise control of stoichiometry. Following changes in the number of conduction electrons by doping may be employed to enhance the value of the critical temperature of BaFe$_2$Se$_3$ itself.